\documentclass[final]{IEEEtran}
\newlength \figwidth
\usepackage{cite}
\ifCLASSINFOpdf
  \usepackage[pdftex]{graphicx}
	\usepackage{epstopdf}
  \graphicspath{{../Figures/}}
  \DeclareGraphicsExtensions{.eps}
\else
  \usepackage[dvips,draft]{graphicx}
\fi
\usepackage[cmex10]{amsmath}
\usepackage{amssymb}
\usepackage{pifont}
\usepackage{amsthm}
\usepackage{url}
\usepackage{dsfont}
\usepackage{tabulary}
\usepackage{multirow}
\usepackage{color}
\newcounter{MYtempeqncnt}

\usepackage{color}
\usepackage{times}
\usepackage{verbatim}
\usepackage{floatflt}
\usepackage{enumerate}
\usepackage{array}
\usepackage{multicol,afterpage,wrapfig} 
\usepackage{tikz}
\usepackage[noend]{algpseudocode}
\makeatletter
\def\BState{\State\hskip-\ALG@thistlm}
\makeatother
\usepackage{amsmath,amssymb,eucal}
\usepackage[utf8]{inputenc}
\usepackage{epsfig}
\usepackage{exscale}
\usepackage{latexsym}
\usepackage{verbatim}
\usepackage{amsfonts}
\usepackage{multirow}
\usepackage{cite}
\usepackage{balance}
\usepackage{color}
\usepackage{transparent}
\usepackage{import}
\usepackage{mathtools}
\usepackage{tikz}
\usetikzlibrary{automata,arrows,positioning,calc}
\usepackage{bbm}
\usepackage[printonlyused,withpage]{acronym}
\usepackage{tabu}
\usepackage{longtable}
\usepackage{balance}
\usepackage{footnote}
\usepackage{tablefootnote}
\def\BibTeX{{\rm B\kern-.05em{\sc i\kern-.025em b}\kern-.08em
    T\kern-.1667em\lower.7ex\hbox{E}\kern-.125emX}}
    
\renewcommand{\IEEEQED}{\IEEEQEDopen}

\newcommand*\xbar[1]{%
  \hbox{%
    \vbox{%
      \hrule height 0.5pt % The actual bar
      \kern0.36ex%         % Distance between bar and symbol
      \hbox{%
        \kern-0.12em%      % Shortening on the left side
        \ensuremath{#1}%
        \kern-0.12em%      % Shortening on the right side
      }%
    }%
  }%
} 

\newcommand{\indic}{\mathds{1}}

\newfont{\bbb}{msbm10 scaled 500}

\newfont{\bb}{msbm10 scaled 1100}

\newcommand{\executeiffilenewer}[3]{%
\ifnum\pdfstrcmp{\pdffilemoddate{#1}}%
{\pdffilemoddate{#2}}>0%
{\immediate\write18{#3}}\fi%
}

\usetikzlibrary{arrows,calc}
\allowdisplaybreaks

\IEEEoverridecommandlockouts
\begin{document}
\pagenumbering{gobble}
%\makeatletter
%\def\thm@space@setup{\thm@preskip=0pt
%\thm@postskip=0pt}
%\makeatother

\newtheorem{Theorem}{\bf Theorem}
\newtheorem{Corollary}{\bf Corollary}
\newtheorem{Remark}{\bf Remark}
\newtheorem{Lemma}{\bf Lemma}
\newtheorem{Proposition}{\bf Proposition}
\newtheorem{Assumption}{\bf Assumption}
\newtheorem{Definition}{\bf Definition}
\title{Operating Massive MIMO in Unlicensed Bands for Enhanced Coexistence and Spatial Reuse}
\author{\IEEEauthorblockN{{Giovanni~Geraci, Adrian~Garcia-Rodriguez, David~L\'{o}pez-P\'{e}rez, Andrea~Bonfante,\\Lorenzo~Galati~Giordano, and Holger~Claussen}}
\thanks{The authors are with Bell Laboratories, Nokia, Dublin, Republic of Ireland (e-mail: dr.giovanni.geraci@gmail.com). The material in this paper will in part be presented at the 2017 IEEE Int. Conf. on Comm. (ICC) \cite{GerGarLop2016ICC}}
}

\maketitle
\thispagestyle{empty}
\begin{abstract}
We propose to operate massive multiple-input multiple output (MIMO) cellular base stations (BSs) in unlicensed bands. We denote such system as massive MIMO unlicensed (mMIMO-U). We design the key procedures required at a cellular BS to guarantee coexistence with nearby Wi-Fi devices operating in the same band. In particular, spatial reuse is enhanced by actively suppressing interference towards neighboring Wi-Fi devices. Wi-Fi interference rejection is also performed during an enhanced listen-before-talk (LBT) phase. These operations enable Wi-Fi devices to access the channel as though no cellular BSs were transmitting, and vice versa. Under concurrent Wi-Fi and BS transmissions, the downlink rates attainable by cellular user equipment (UEs) are degraded by the Wi-Fi-generated interference. To mitigate this effect, we select a suitable set of UEs to be served in the unlicensed band accounting for a measure of the Wi-Fi/UE proximity. Our results show that the so-designed mMIMO-U allows simultaneous cellular and Wi-Fi transmissions by keeping their mutual interference below the regulatory threshold. Compared to a system without interference suppression, Wi-Fi devices enjoy a median interference power reduction of between 3~dB with 16 antennas and 18~dB with 128 antennas. With mMIMO-U, cellular BSs can also achieve large data rates without significantly degrading the performance of Wi-Fi networks deployed within their coverage area.
\end{abstract}
\IEEEpeerreviewmaketitle
\section{Introduction}
%\cite{YanGerQueTSP2016}
In view of the ever increasing mobile data demand, the wireless industry has turned its attention to unlicensed spectrum bands, e.g., 5~GHz, to provide extra frequency resources for the fifth generation (5G) cellular networks~\cite{Huawei2013,QualcommMulteFire2015,NokiaMulteFire2015,ZhaWanCai2015}. In 5G communication systems, licensed-unlicensed integration may allow mobile operators to serve more users via traffic offloading and/or to enhance their peak data rate through carrier aggregation. Besides, standalone unlicensed technologies may unlock new vertical markets and their corresponding revenues.
On the other hand, harmonious coexistence with other technologies working in the unlicensed spectrum, such as IEEE~802.11x (Wi-Fi), must be guaranteed~\cite{ZhaChuGuo2015,MukCheFal2015,BenSimCzy2013}. This is because Wi-Fi systems rely on a contention-based access with a random backoff mechanism,
i.e., carrier sensing multiple access/collision avoidance (CSMA-CA)~\cite{PerSta2013}. Therefore, cellular base stations (BSs) transmitting continuously over unlicensed bands would produce harmful interference and generate repeated backoffs at the Wi-Fi nodes.

\subsection{Background and Motivation}

Two main approaches are currently under consideration by network operators to exploit the unlicensed band and guarantee coexistence between cellular BSs and Wi-Fi devices. 
Both augment an existing licensed band interface with supplemental unlicensed band downlink transmissions.

\subsubsection{Long Term Evolution unlicensed (LTE-U)}

LTE-U uses carrier-sensing adaptive transmission (CSAT) and it is mainly targeted at the United States market, 
where channel sensing operations prior to transmission are not required~\cite{LTEUForum:15}. 
With CSAT, cellular BSs interleave their transmissions with idle intervals,
which allow Wi-Fi devices to access the channel~\cite{RahBehKoo2011,Qualcomm2014}. 
For example, a cellular BS may access the channel at every other frame boundary, i.e., transmitting for a 10ms frame, 
then leaving the channel idle for the next 10ms frame, etc., 
thus yielding a 50\% on-off duty cycle. 
As a result, every channel use gained by the cellular BS comes at the expense of idle periods at the Wi-Fi devices.

\subsubsection{Licensed Assisted Access (LAA)}

In LAA, cellular BSs sense the channel activity via energy detection, 
and they commence a transmission in the unlicensed band only if the channel is deemed free for a designated period of time~\cite{3GPP36889,RatManGho2014}. 
Such channel sensing operation, denoted as listen before talk (LBT), is mandatory in some regions, 
e.g., Europe and Japan~\cite{3GPP-RP-140808,Nokia:14}. 
Similarly to the random access procedure used by Wi-Fi devices, 
LBT employs random backoff intervals and a variable exponentially distributed contention window size. 
The latter is recommended by the 3GPP as the baseline approach for downlink transmissions to guarantee a fair sharing of time resources with Wi-Fi devices~\cite{CanLopCla2016}.

While ensuring coexistence, 
both LTE-U/CSAT and LAA/LBT are based on discontinuous transmission, 
i.e., neither allows simultaneous usage of the unlicensed spectrum by both cellular BSs and Wi-Fi devices when their coverage areas overlap. 
This may be a conservative approach in certain scenarios, 
mostly when multiple antennas are available. 
In fact, multiple antennas could be used by cellular BSs to increase spatial reuse and provide additional throughput without diminishing the Wi-Fi data rates.

\begin{figure*}[!t]
\centering
\includegraphics[width=\figwidth]{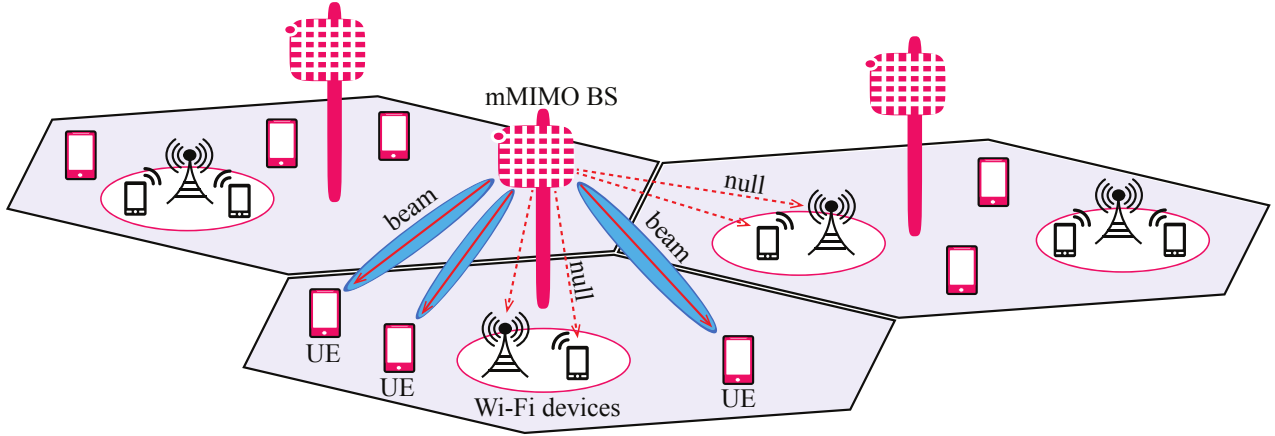}
\caption{Illustration of a mMIMO-U system: Each BS multiplexes UEs in the unlicensed band while suppressing interference at neighboring Wi-Fi devices.}
\label{fig:SystemModel}
\end{figure*}

\subsection{Approach and Contributions}

We propose massive multiple-input multiple-output (MIMO) as a means to enhance coexistence, 
while maximizing spectrum reuse in the unlicensed band. 
Massive MIMO has recently emerged as one of the potential disruptive technologies for the 5G wireless systems, 
where cellular BSs are envisioned to be equipped with a large number of antennas~\cite{Mar:10,RusPerBuoLar:2013,LuLiSwi2014,LarEdfTuf2014,BjoLarDeb2016}. 
In this paper, we consider a downlink massive MIMO system operating in the unlicensed band. 
We refer to this system as massive MIMO unlicensed (mMIMO-U). 
In the proposed system, 
a subset of the spatial degrees of freedom (d.o.f.) provided by the large number of antennas are employed to suppress the mutual interference between each massive MIMO BS and the Wi-Fi devices operating in its neighborhood. 
This allows massive MIMO BSs and Wi-Fi devices to access unlicensed bands simultaneously, 
thus increasing the network spatial reuse.
The remaining spatial d.o.f. are used by the massive MIMO BS to multiplex multiple data streams. 

The present work is expected to advance the understanding of 5G cellular networks operating in the unlicensed spectrum, 
where newly deployed massive MIMO and existing Wi-Fi systems may coexist. 
On the basis of the key principles of both technologies, 
we identify the rich research opportunities and tackle the fundamental challenges that arise when operating massive MIMO in the unlicensed band. 
Our contributions can be summarized as follows.
\begin{itemize}
\item \textit{Scheduling:} 
We discuss the operations required for a massive MIMO cellular BS to: 
\emph{(i)} acquire channel state information from the neighboring Wi-Fi devices, 
\emph{(ii)} allocate spatial resources for Wi-Fi interference suppression and user equipment (UE) multiplexing, 
and \emph{(iii)} select a suitable set of UEs to be served in the unlicensed band.% Such operations are distributed and do not entail inter-cell coordination.
\item \textit{Transmission:} 
We devise the key transmission operations of a mMIMO-U system, including 
\emph{(i)} an enhanced LBT phase, 
\emph{(ii)} procedures for UE pilot request and channel estimation, 
and \emph{(iii)} precoder calculation. 
In all of the above phases, 
the large number of BS antennas is exploited to suppress interference to/from neighboring Wi-Fi devices, 
so that cellular/Wi-Fi coexistence is improved.
\item \textit{Performance:} 
We evaluate the performance of the proposed mMIMO-U operations in scenarios of practical interest. 
We show that mMIMO-U significantly reduces mutual interference between massive MIMO cellular BS and Wi-Fi devices,
while multiplexing a number of data streams. 
As a result, large cellular data rates can be achieved without significantly degrading the performance of Wi-Fi networks deployed within the coverage area of a massive MIMO cellular BS.
\end{itemize}

%\subsubsection*{Paper outline} We introduce the mMIMO-U system model in Section~II. In Section~III, we discuss UE scheduling and spatial resource allocation for interference suppression. In Section~IV, we design an enhanced LBT phase, and pilot and data transmission operations for the unlicensed band. In Section~V, we provide simulation results to evaluate the performance of the proposed mMIMO-U. We conclude the paper in Section~VI.

\subsubsection*{Notations} Capital and lower-case bold letters denote matrices and vectors, respectively. The superscripts $[\mathbf{X}]^*$, $[\mathbf{X}]^\mathrm{T}$, and $[\mathbf{X}]^\mathrm{H}$ denote conjugate, transpose, and conjugate transpose, respectively. The notation $\widehat{\mathbf{X}}$ denotes an approximation or estimate of $\mathbf{X}$. The subspace spanned by the columns of $\mathbf{X}$ and its orthogonal subspace are denoted $\mathrm{range}\{\mathbf{X}\}$ and $\mathrm{null}\{\mathbf{X}\}$, respectively. Given a set $\mathcal{X}$, $\mathrm{card}\{\mathcal{X}\}$ denotes its cardinality. 
\section{System Set-Up}

We now provide a general introduction to the network topology and channel model used in this paper. 
More details on the specific parameters used for our numerical studies will be given in Section~\ref{sec:simulations}.

\subsection{Network Topology}

We consider the downlink of a cellular network, as shown in Fig.~\ref{fig:SystemModel}, where massive MIMO cellular BSs are deployed to operate in the unlicensed band in a synchronous manner, and communicate with their respective sets of connected cellular UEs, 
while multiple Wi-Fi devices, i.e., access points (APs) and stations (STAs), also operate in the same unlicensed band. 
It is important to note that even if cellular BSs may also have access to a licensed band, 
this paper will focus on cellular BS operations and transmissions in the unlicensed band. 

On the cellular side, 
we denote by $\mathcal{I}$  the sets of cellular BSs,
and assume that all cellular BSs transmit with power $P_{\textrm{b}}$.
Cellular UEs associate to the cellular BS that provides the largest average received power. 
Each BS $i$ is equipped with a large number of antennas $N$, 
and it simultaneously serves $K_i$ of its associated UEs, $K_i \leq N$, on each time-frequency resource block (RB) through spatial multiplexing. 
Each UE has a single antenna. 
We denote $\mathcal{K}_i$ the set of UEs served by BS $i$ in the unlicensed band. 
It is important to note that while the total number of associated UEs is determined by the UE density and distribution as well as  the nature of traffic, 
the value of $K_i$ can be chosen adaptively by BS $i$ via scheduling operations. 

\begin{figure*}[!t]
\normalsize
\setcounter{MYtempeqncnt}{\value{equation}}
\hrulefill
\begin{align}
y_{ik}[m] = \!\sqrt{P_{\textrm{b}}} {\mathbf{h}_{iik}^{\mathrm{H}} \mathbf{w}_{ik} s_{ik}[m]} + \!\sqrt{P_{\textrm{b}}} \!\!\!\! \sum_{k'\in\mathcal{K}_i\backslash k} \!\!\!\!\!  {\mathbf{h}_{iik}^{\mathrm{H}} \mathbf{w}_{ik'} s_{ik'}[m]}
+ \! \sqrt{P_{\textrm{b}}} \! \sum_{i' \in \mathcal{I} \backslash i} \,\sum_{\,k'\in\mathcal{K}_{i'}} \!\! {\mathbf{h}_{i'ik}^{\mathrm{H}} \mathbf{w}_{i'k'} s_{i'k'}[m]} \!+\! \! \sum_{\ell \in \mathcal{L}^{\mathrm{a}}} \!\! \sqrt{P_{\ell}} \, {q_{\ell ik} s_{\ell}[m]} \!+\! \epsilon_{ik}[m]
\label{eqn:rx_signal}
\end{align}
%\hrulefill
\setcounter{equation}{\value{MYtempeqncnt}}
\addtocounter{equation}{1}
\setcounter{MYtempeqncnt}{\value{equation}}
\begin{align}\label{equ:sinr_ue}
\nu_{ik} = \frac{ P_{\textrm{b}} \vert \mathbf{h}_{iik}^{\mathrm{H}} \mathbf{w}_{ik} \vert^2 }
{P_{\textrm{b}} \sum_{k'\in\mathcal{K}_i\backslash k} \vert \mathbf{h}_{iik}^{\mathrm{H}} \mathbf{w}_{ik'} \vert^2 + 
P_{\textrm{b}} \sum_{i' \in \mathcal{I} \backslash i} \sum_{k'\in\mathcal{K}_{i'}} \vert \mathbf{h}_{i'ik}^{\mathrm{H}} \mathbf{w}_{i'k'} \vert^2 + 
\sum_{\ell \in \mathcal{L}^{\mathrm{a}}} P_{\ell} \vert q_{\ell ik} \vert^2
+ \sigma^2_{\epsilon}}
\end{align}
\hrulefill
\setcounter{equation}{\value{MYtempeqncnt}}
\vspace*{-3pt}
\end{figure*}
\addtocounter{equation}{1}

On the Wi-Fi side,
we denote by $\mathcal{L}_{\textrm{AP}}$, $\mathcal{L}_{\textrm{STA}}$, and $\mathcal{L}$ the sets of Wi-Fi APs, Wi-Fi stations, and all Wi-Fi devices, respectively, 
which are assumed to be invariant. 
Moreover, we denote by  $\mathcal{L}_{\textrm{AP}}^{\mathrm{a}}$, $\mathcal{L}_{\textrm{STA}}^{\mathrm{a}}$, and $\mathcal{L}^{\mathrm{a}}$ the sets of active Wi-Fi APs, stations, and all Wi-Fi devices, respectively,
which vary according to the Wi-Fi traffic profile~\cite{PerSta2013},
and assume that all Wi-Fi devices $\ell$ transmit with power $P_{\ell}$.
STAs associate to the Wi-Fi AP that provides the largest average received power. 
Each AP is equipped with a single antenna,
and serves one STA at the time. 
Each STA has a single antenna.

\subsection{Channel Model}

All propagation channels are affected by slow fading (comprising antenna gain, path loss, and shadowing) and fast fading, 
as detailed in Section~\ref{sec:simulations}. 
We adopt a block-fading propagation model, 
where the propagation channels are assumed constant within their respective time/frequency coherence intervals~\cite{MedTse2000}. 

Without loss of generality, 
and assuming a single antenna for all UEs and Wi-Fi devices, 
we define the following variables for a given time/frequency coherence interval:
\begin{itemize}
\item 
$\mathbf{h}_{ijk} = \sqrt{\bar{h}_{ijk}} \tilde{\mathbf{h}}_{ijk} \in \mathbb{C}^{N\times 1}$ 
denotes the channel vector between BS $i$ and UE $k$ in cell $j$;
\item 
$\mathbf{g}_{i\ell} = \sqrt{\bar{g}_{i\ell}} \tilde{\mathbf{g}}_{i\ell} \in \mathbb{C}^{N\times 1}$ 
denotes the channel vector between BS $i$ and Wi-Fi device $\ell$;
\item 
$q_{\ell jk} = \sqrt{\bar{q}_{\ell jk}} \tilde{q}_{\ell jk} \in \mathbb{C}$ 
denotes the channel coefficient between Wi-Fi device $\ell$ and UE $k$ in cell $j$.
\end{itemize}
In the above, 
the coefficients $\bar{h}_{ijk} \in \mathbb{R}^+$, $\bar{g}_{i\ell} \in \mathbb{R}^+$, and $\bar{q}_{\ell jk} \in \mathbb{R}^+$ represent the respective slow fading gains, 
which are assumed constant. 
The coefficients $\tilde{\mathbf{h}}_{ijk} \in \mathbb{C}^{N\times 1}$, $\tilde{\mathbf{g}}_{i\ell} \in \mathbb{C}^{N\times 1}$, and $\tilde{q}_{\ell jk} \in \mathbb{C}$ contain the respective fast fading, which varies at every time/frequency coherence interval.

Without loss of generality, 
we also assume the same symbol duration for cellular and Wi-Fi transmissions. 
Thus, the signal $y_{ik}[m] \in \mathbb{C}$ received by UE $k$ in cell $i$ at symbol interval $m$ can be expressed as (\ref{eqn:rx_signal}) at the top of this page, 
where 
\emph{(i)} $\mathbf{w}_{ik} \in \mathbb{C}^{N\times 1}$ is the precoding vector from BS $i$ to UE $k$ in cell $i$, 
\emph{(ii)} $s_{ik}[m] \in \mathbb{C}$ is the unit-variance signal intended for UE $k$ in cell $i$, 
\emph{(iii)} $s_{\ell}[m] \in \mathbb{C}$ is the unit-variance signal transmitted by Wi-Fi device $\ell$, 
and \emph{(iv)} $\epsilon_{ik}[m] \sim \mathcal{CN}(0,\sigma^2_{\epsilon})$ is the thermal noise. 
The five terms on the right hand side of (\ref{eqn:rx_signal}) respectively represent: 
useful signal, intra-cell interference from the serving BS, inter-cell interference from other BSs, interference from Wi-Fi devices, and thermal noise. 

The resulting signal to interference plus noise ratio (SINR) $\nu_{ik}$ at UE $k$ in cell $i$ is obtained via an expectation over all symbols, 
and it is given by (\ref{equ:sinr_ue}) at the top of this page.

The corresponding interference power $I_{: \rightarrow \ell}[m]$ received at Wi-Fi device $\ell$ due to cellular downlink operations is given by
\begin{equation}
I_{: \rightarrow \ell}[m] = P_{\textrm{b}} \sum_{i\in \mathcal{I}} \sum_{k\in\mathcal{K}_i} \left| \mathbf{g}_{il}^{\mathrm{H}} \mathbf{w}_{ik} s_{ik}[m] \right|^2.
\label{eqn:Il}
\end{equation}
Each Wi-Fi device deems the channel as occupied and defers from transmission when the total received power, 
i.e., from all cellular BSs and all other Wi-Fi devices,
falls above the regulatory threshold $\gamma_{\mathrm{LBT}}$.
\section{Massive MIMO Unlicensed Scheduling}
\label{sec:scheduling}

\begin{figure*}[!t]
\centering
\includegraphics[width=\figwidth]{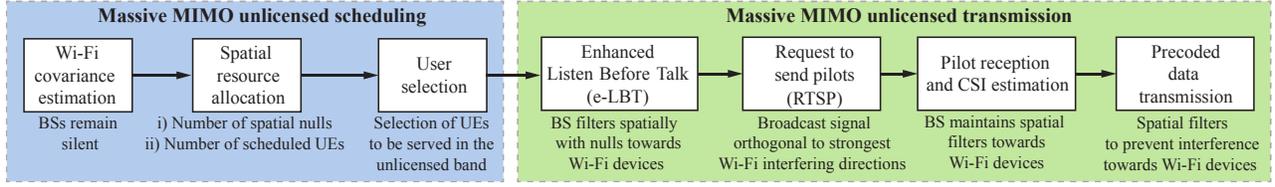}
\caption{Flow chart of the proposed mMIMO-U procedures: Operations to the left (resp. right) are detailed in Section~\ref{sec:scheduling} (resp. Section~\ref{sec:operations}).}
\label{fig:FlowChart}
\end{figure*}

In this section, we discuss scheduling operations for the proposed mMIMO-U system. We first detail the necessary procedures for a BS to acquire channel state information (CSI) from the neighboring Wi-Fi devices. 
Then we discuss the spatial resource allocation, 
i.e., how to choose the number of spatial d.o.f. to be allocated for Wi-Fi interference suppression and for UE multiplexing, respectively. 
Finally, we devise a UE selection scheme for choosing the UEs to be served in the unlicensed band. The sequence of operations presented in this section takes place at every Wi-Fi channel coherence interval, 
and is outlined in the three leftmost blocks of Fig.~\ref{fig:FlowChart}. %$C_{\mathrm{c}}$

\subsection{Wi-Fi Channel Covariance Estimation}
\label{sec:covariance_estimation}

In order to suppress interference to/from Wi-Fi devices, each BS $i$ requires information about the BS-to-Wi-Fi channels. In our proposed mMIMO-U system, BS $i$ periodically obtains the channel subspace occupied by neighboring Wi-Fi devices through a channel covariance estimation procedure, presented in the following.

Throughout the channel covariance estimation procedure, all BSs remain silent, and thus each BS $i$ receives the signal
\begin{equation}
\mathbf{z}_i[m] = \sum_{\ell\in \mathcal{L}^{\mathrm{a}}} \sqrt{P_{\ell}} \mathbf{g}_{i\ell} s_{\ell}[m] + \boldsymbol{\eta}_i[m],
\label{eqn:zi}
\end{equation}
which consists of all transmissions from active Wi-Fi devices and a noise term $\boldsymbol{\eta}_i[m] \sim \mathcal{CN}(\mathbf{0},\sigma^2_{\eta} \mathbf{I})$.\footnote{In the case of other cellular operators using the same unlicensed band, (\ref{eqn:zi}) would include their transmitted signals, and the mMIMO-U operations would ensure coexistence with these operators as well as with Wi-Fi devices.}

Let us now denote by $\mathbf{Z}_i \in \mathbb{C}^{N \times N}$ the covariance of $\mathbf{z}_i[m]$, 
which can be defined as
\begin{equation}
\mathbf{Z}_i =  \mathbb{E} \left[ \mathbf{z}_i \mathbf{z}_i^{\mathrm{H}}\right],
\label{eqn:Z}
\end{equation}
where the expectation is taken with respect to $\mathcal{L}^{\mathrm{a}}$, $s$, and $\boldsymbol{\eta}$. Then, BS $i$ can obtain an estimate $\widehat{\mathbf{Z}}_i$ of $\mathbf{Z}_i$ via a simple average over $M_{\mathrm{c}}$ symbol intervals as~\cite{HoyHosTen2014}
\begin{equation}
\widehat{\mathbf{Z}}_i = \frac{1}{M_{\mathrm{c}}} \sum_{m=1}^{M_{\mathrm{c}}} \mathbf{z}_i[m] \mathbf{z}_i^{\mathrm{H}}[m],
\label{eqn:Z_hat}
\end{equation}
where the value of $M_{\mathrm{c}}$ must be sufficiently large to ensure that all neighboring Wi-Fi devices were active. Other possible approaches to channel covariance matrix estimation are discussed in \cite{BjoSanDeb2016} and references therein. It is obvious that the operation in (\ref{eqn:Z_hat}) incurs an overhead, and an inherent trade-off exists between improving the quality of the estimate in (\ref{eqn:Z_hat}) and reducing the overhead. 

Given the estimate $\widehat{\mathbf{Z}}_i$, BS $i$ applies a spectral decomposition, obtaining
\begin{equation}
\widehat{\mathbf{Z}}_i = \widehat{\mathbf{U}}_i \widehat{\mathbf{\Lambda}}_i \widehat{\mathbf{U}}_i^{\mathrm{H}},
\label{eqn:spectral}
\end{equation}
where the columns of $\widehat{\mathbf{U}}_i=[\widehat{\mathbf{u}}_{i1},\ldots,\widehat{\mathbf{u}}_{iN}]$ form an orthonormal basis, 
and
\begin{equation}
\widehat{\mathbf{\Lambda}}_i = \textrm{diag}\left( \widehat{\lambda}_{i1},\ldots,\widehat{\lambda}_{iN} \right)	
\end{equation}
contains a set of eigenvalues, 
such that $\widehat{\lambda}_{i1} \geq \widehat{\lambda}_{i2} \ldots \geq \widehat{\lambda}_{iN}$.
Then, with the proposed mMIMO-U operations, 
BS $i$ can allocate a certain number of spatial d.o.f., denoted as $D_i$, to suppress interference to/from the dominant directions of the Wi-Fi channel subspace. As it will be shown in Section~\ref{sec:simulations}, a sufficiently large value of $D_i$ is required in order to ensure coexistence. Note that a different unlicensed frequency channel can be selected when the presence of a Wi-Fi device is detected too close to the cellular BS.

In order to allow such d.o.f. allocation procedure, let us now define the matrix
\begin{equation}
\widehat{\mathbf{\Sigma}}_i \triangleq \left[ \widehat{\mathbf{u}}_{i1},\ldots,\widehat{\mathbf{u}}_{iD_i} \right],	
\label{eqn:Sigma}
\end{equation}
whose columns contain the $D_i$ dominant eigenvectors of $\widehat{\mathbf{Z}}_i$. For a sufficiently large $D_i$, $\mathrm{range}\{\widehat{\mathbf{\Sigma}}_i\}$ represents the channel subspace on which BS $i$ receives most of the Wi-Fi-transmitted power. Since Wi-Fi uplink/downlink and BS downlink transmissions share the same frequency band, channel reciprocity holds. Therefore, the power transmitted by BS $i$ on $\mathrm{range}\{\widehat{\mathbf{\Sigma}}_i\}$ represents the major source of interference for one or more Wi-Fi devices.

\subsection{Spatial Resource Allocation}
\label{sec:resource_allocation}

From the Wi-Fi covariance estimate in (\ref{eqn:Z_hat}), each BS performs spatial resource allocation. In particular, each BS calculates the number of UEs $K_i$ to spatially multiplex in the unlicensed band and the number of spatial d.o.f. $D_i$ to be allocated to suppress interference to/from neighboring Wi-Fi devices. 
To this end, a variety of criteria can be employed to select the pair $(K_i,D_i)$, which must satisfy
\begin{equation}
D_{i} \leq \min \left\{ N - K_i, M_{\mathrm{c}} \right\}.
\label{eqn:Di_fixed_Ki}
\end{equation}
The inequality in (\ref{eqn:Di_fixed_Ki}) indicates that $D_i+K_i$ should not exceed the available spatial d.o.f. $N$ at the BS, and that $D_i$ should be upper bounded by the rank $M_{\mathrm{c}}$ of $\widehat{\mathbf{Z}}_i$ defined in (\ref{eqn:Z_hat}).\footnote{Note that for sparse communication channels, the number of available spatial d.o.f. may be lower than the number of antennas $N$ \cite{ShiFosGan2000}.} Intuitively, the value of $D_i$ controls the number of excess d.o.f. used for interference suppression, and it is chosen by compromising beamforming gain at the UEs for enhanced interference suppression to/from Wi-Fi devices. For instance, $D_i$ could be set as the number of dominant eigenvectors of $\widehat{\mathbf{Z}}_i$, i.e., those containing a given percentage of received Wi-Fi signal power. We refer the reader to \cite{BjoLarDeb2016} for a relevant study on the choice of $K_i$.

\subsection{UE Selection}
\label{sec:user_selection}

Power emissions in the unlicensed band are strictly regulated. In some countries, the maximum allowed transmit power decreases with the number of antenna elements, if these are used to focus energy in a particular direction~\cite{FCC2013}. 
This means that the coverage area of mMIMO-U BSs may be limited, and only UEs sufficiently close to the BS should be scheduled in the unlicensed band. 
Moreover, when cellular BSs and Wi-Fi devices simultaneously operate in the unlicensed band, 
the Wi-Fi-to-UE interference may degrade the cellular downlink rates. Therefore, a mMIMO-U BS should select UEs that are not in the proximity of a Wi-Fi device.

In light of the above, and based on information available from protocols currently implemented, 
we propose for each BS $i$ to rank the associated cellular UEs according to the following metric
\begin{equation}
\mu_{ik} = \frac{ P_{\textrm{b}} \bar{\mathbf{h}}_{iik}}
{P_{\textrm{b}} \sum_{i' \in \mathcal{I} \backslash i} \bar{\mathbf{h}}_{i'ik} + \sum_{\ell\in \mathcal{L}_{\textrm{AP}}} P_{\ell} \bar{q}_{\ell ik}}.
\label{eqn:scheduler_metric}
\end{equation}
Intuitively, $\mu_{ik}$ represents a measure of the average SINR received at UE $k$ in cell $i$ during a non-precoded broadcast signal transmission. 
The metric in (\ref{eqn:scheduler_metric}) accounts for the average channel gain between the UE and: the serving BS, other BSs, and Wi-Fi APs. 
In practical implementations, 
$\mu_{ik}$ can be obtained via the two following steps.
\begin{itemize}
\item 
An accurate estimation of $\bar{\mathbf{h}}_{iik}$ can be obtained at the UE through downlink measurements on the cell reference signal (CRS), 
by subtracting the reference signal power (signaled by the BS) from the reference signal received power (RSRP)~\cite{3GPP36201,3GPP36213}. 
The fast fading component $\tilde{\mathbf{h}}_{iik}$ is removed by filtering the measurements over a time window~\cite{ZhaSolLia2012}. 
Applying the same procedure on the CRS transmitted by other BSs yields the terms $\bar{\mathbf{h}}_{i'ik}, \, i' \in \mathcal{I} \backslash i$.
%\cite{3GPP36331}. 
\item 
The value of $\bar{q}_{\ell ik}$ can be obtained through the automatic neighbor relations (ANR) function, 
where BS $i$ requests UE $k$ to report Wi-Fi measurements that contain the received signal strength indicator (RSSI) from Wi-Fi AP $\ell$~\cite{3GPP36300}. 
Typically, few dominant terms $\bar{q}_{\ell ik}$ are sufficient to estimate $\mu_{ik}$, 
as the UE is unlikely to be close to a multitude of Wi-Fi APs at the same time.
\end{itemize}
The above measurements can be fed back by the UE to the BS on a licensed control signaling interface~\cite{CanLopCla2016}. The BS then selects the $K_i$ UEs with the highest metric $\mu_{ik}$ for transmission in the unlicensed band. UEs that are not selected, e.g., because they are co-located with a WiFi device, may be scheduled for transmission in the licensed band, or may wait to be re-scheduled when their channel conditions have varied, e.g., because the UE or Wi-Fi are not transmitting or their positions have changed.

The advantage of using the proposed metric $\mu_{ik}$ instead of instantaneous CSI for scheduling purposes lies in the fact that $\mu_{ik}$ varies on a slow scale. Thus, feedback from the UEs does not need to be requested at every BS-UE channel coherence interval, and the resulting overhead is lower.
\section{Massive MIMO Unlicensed Transmission}
\label{sec:operations}

The main operations we propose to perform at the mMIMO-U BSs for data transmission are: enhanced LBT, UE pilot request and channel estimation, and precoder calculation, as outlined in the four rightmost blocks of Fig.~\ref{fig:FlowChart}. In all of the above operations, the large number of transmit antennas available at the BSs is exploited to suppress interference to/from neighboring Wi-Fi devices, so that both cellular BSs and Wi-Fi devices can simultaneously use the unlicensed band.

\subsection{Enhanced Listen Before Talk}
\label{sec:eLBT}

In order to comply with the regulations in the unlicensed band, 
each BS must perform LBT before any data transmission~\cite{3GPP-RP-140808}. In current coexistence approaches, such as LAA, 
a transmission opportunity is gained by BS $i$ if the sum power received from all devices using the same band falls below the regulatory threshold for a designated time interval, i.e., 
\begin{equation}
I_{i \leftarrow :}[m] = \| \mathbf{z}_i[m]\|^2 < \gamma_{\mathrm{LBT}}, \enspace m=1,\ldots,M_{\mathrm{LBT}},
\label{eqn:LBT_conventional}
\end{equation}
where $\mathbf{z}_i[m]$ is given in (\ref{eqn:zi}), 
and the duration $M_{\mathrm{LBT}}$ is given by a distributed inter-frame space (DIFS) interval plus a random number of backoff time slots~\cite{PerSta2013}. The process in (\ref{eqn:LBT_conventional}) is also known as energy detection. LBT may be sometimes conservative allowing for the transmission of either a single BS or a Wi-Fi device within a certain coverage area, thus preventing spatial reuse of the same unlicensed band.

In the proposed mMIMO-U system, the LBT phase is enhanced as follows. When BS $i$ listens to the transmissions currently taking place in the unlicensed band, it filters the received signal $\mathbf{z}_i[m]$ with the $D_i$ spatial nulls defined in (\ref{eqn:Sigma}). Let us define the following matrices
\begin{equation}
\widehat{\mathbf{\Pi}}_i = \widehat{\mathbf{\Sigma}}_i \widehat{\mathbf{\Sigma}}_i^{\mathrm{H}},
\label{eqn:Pi}
\end{equation}
which projects a vector onto the subspace $\mathrm{range}\{\widehat{\mathbf{\Sigma}}_i\}$, and
\begin{equation}
\widehat{\mathbf{\Pi}}_i^{\perp} = \mathbf{I} - \widehat{\mathbf{\Pi}}_i = \mathbf{I} - \widehat{\mathbf{\Sigma}}_i \widehat{\mathbf{\Sigma}}_i^{\mathrm{H}},
\label{eqn:Pi_perp}
\end{equation}
which projects a vector onto $\mathrm{null}\{\widehat{\mathbf{\Sigma}}_i\}$. 
A transmission opportunity is then gained by BS $i$ if the condition
\begin{equation}
I_{i \leftarrow :}[m] = \left\| \widehat{\mathbf{\Pi}}_i^{\perp} \mathbf{z}_i[m] \right\|^2 < \gamma_{\mathrm{LBT}}, \enspace m=1,\ldots,M_{\mathrm{LBT}},
\label{eqn:LBT}
\end{equation}
holds for $M_{\mathrm{LBT}}$ symbols. 

In other words, since the channel subspace $\mathrm{range}\{\widehat{\mathbf{\Sigma}}_i\}$ is occupied by neighboring Wi-Fi devices, BS $i$ may transmit downlink signals on the channel subspace $\mathrm{null}\{\widehat{\mathbf{\Sigma}}_i\}$ only, and it must ensure that no concurrent transmissions are detected on $\mathrm{null}\{\widehat{\mathbf{\Sigma}}_i\}$. This is accomplished by measuring the aggregate power of the received signal filtered through the projection $\widehat{\mathbf{\Pi}}_i^{\perp}$. Provided that a sufficient number of d.o.f. $D_i$ have been allocated for interference suppression, the condition in (\ref{eqn:LBT}) is met. Therefore, unlike conventional LBT operations, the enhanced LBT (e-LBT) phase allows both mMIMO-U BSs and Wi-Fi devices to simultaneously access the unlicensed band.

\subsection{UE Channel Estimation}
\label{sec:channel_estimation}

\begin{figure}[!t]
\centering
\includegraphics[width=0.85\columnwidth]{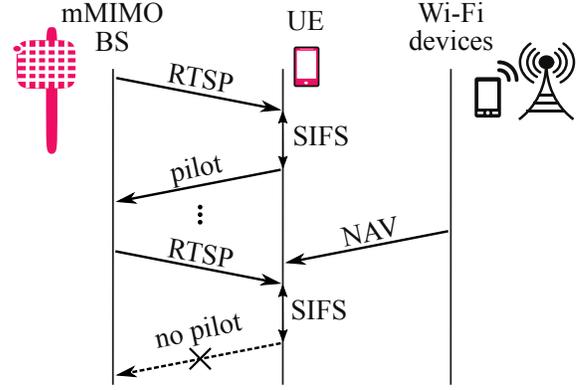}
\caption{A UE responds to RTSP messages with omnidirectional pilot signals, unless a network allocation vector message is received from a Wi-Fi device.}
\label{fig:RTSP_NAV}
\end{figure}

Once the LBT procedure has succeeded, in order to spatially multiplex the $K_i$ selected UEs, BS $i$ requires knowledge of their channels $\mathbf{h}_{iik}$, $k\in\mathcal{K}_i$. UE CSI may be obtained via pilot signals transmitted during a training phase at every BS-UE channel coherence interval, where coexistence between uplink pilots sent by UEs and Wi-Fi transmissions must be guaranteed. In the proposed UE channel estimation phase, BS $i$ addresses the selected UEs with a request to send pilots (RTSP) message, as shown in Fig.~\ref{fig:RTSP_NAV}. The RTSP message is transmitted on the subspace $\mathrm{null}\{\widehat{\mathbf{\Sigma}}_i\}$, such that interference generated at neighboring Wi-Fi devices is suppressed.\footnote{UEs may receive RTSP messages superimposed with concurrent Wi-Fi transmissions. However, this is unlikely to occur thanks to the UE selection metric in (\ref{eqn:scheduler_metric}), which tends to schedule UEs that are far from Wi-Fi APs.} The addressed UEs respond by simultaneously transmitting back omnidirectional pilot signals after a short inter-frame space (SIFS) time interval~\cite{PerSta2013}.\footnote{The UE selection metric in (\ref{eqn:scheduler_metric}) also ensures that pilots do no create significant interference at Wi-Fi devices. Moreover, a UE may be informed of ongoing nearby Wi-Fi transmissions via network allocation vector (NAV) messages and thus decide to refrain from transmitting its pilot~\cite{PerSta2013}.}

%with $1 \leq M_{\mathrm{p}} \leq C_{\mathrm{p}}$
Let the pilot signals span $M_{\mathrm{p}}$ symbols on each coherence interval. The pilot transmitted by UE $k$ in cell $i$ is denoted by $\mathbf{v}_{\textrm{i}_{ik}} \in \mathbb{C}^{M_{\mathrm{p}}}$, where $\textrm{i}_{ik}$ is the index in the pilot codebook, and all pilots form an orthonormal basis \cite{JosAshMar2011,NgoLarMar2013}. Each pilot signal received at the BS suffers contamination due to pilot reuse across mMIMO-U cells and due to concurrent Wi-Fi transmissions. 
The collective received signal at BS $i$ is denoted as $\mathbf{Y}_i \in \mathbb{C}^{N \times M_{\mathrm{p}}}$ and given by
\begin{equation}
\mathbf{Y}_i = \sum_{j \in \mathcal{I}} \sum_{k\in\mathcal{K}_j} \sqrt{P_{jk}} \mathbf{h}_{ijk} \mathbf{v}_{\mathrm{i}_{jk}}^{\textrm{T}} +
\sum_{\ell \in \mathcal{L}^{\mathrm{a}}} \sqrt{P_{\ell}} {\mathbf{g}_{i \ell} \mathbf{s}_{\ell}^{\mathrm{T}}} + \mathbf{N}_i,	
\label{eqn:Yi}
\end{equation}
where $\mathbf{s}_{\ell} = [s_{\ell}[1],\ldots,s_{\ell}[M_{\mathrm{p}}]]$, $\mathbf{N}_i$ contains additive noise at BS $i$ during pilot signaling, and $P_{jk}$ is the power transmitted by UE $k$ in cell $j$. We assume fractional uplink power control as follows \cite{UbeVilRos2008,R1073224}
\begin{equation}
P_{jk} = \min\left\{ P_{\textrm{max}}, P_0 \cdot \bar{h}_{jjk}^\alpha \right\},
\label{eqn:power_control}
\end{equation}
where $P_{\textrm{max}}$ is the maximum UE transmit power, 
$P_0$ is a cell-specific parameter, $\alpha$ is a path loss compensation factor, and $\bar{h}_{jjk}$ is the slow fading measured at UE $k$ in cell $j$ based on the RSRP~\cite{3GPP36201,3GPP36213}. The aim of (\ref{eqn:power_control}) is to compensate only for a fraction $\alpha$ of the path loss, 
up to a limit imposed by $P_{\textrm{max}}$.

The received signal $\mathbf{Y}_i$ in (\ref{eqn:Yi}) is processed at BS $i$ by \emph{(i)} correlating it with the known pilot signal $\mathbf{v}_{\textrm{i}_{ik}}$, and \emph{(ii)} projecting it onto $\mathrm{null}\{\widehat{\mathbf{\Sigma}}_i\}$. The above operations respectively reject interference from \emph{(i)} orthogonal pilots, and \emph{(ii)} neighboring Wi-Fi transmissions. BS $i$ thus obtains the following CSI estimate for UE $k$ in cell $i$~\cite{kay}
\begin{equation}
\begin{aligned}
\widehat{\mathbf{h}}_{iik} &=
 \widehat{\mathbf{\Pi}}_i^{\perp} \mathbf{Y}_i \mathbf{v}_{\mathrm{i}_{ik}}^{*} = \sqrt{P_{ik}} \widehat{\mathbf{\Pi}}_i^{\perp} \mathbf{h}_{iik} \\
&+ \widehat{\mathbf{\Pi}}_i^{\perp} \Big( \sum_{j \in \mathcal{I} \backslash i} \,\sum_{k\in\mathcal{K}_j} \sqrt{P_{jk}} \mathbf{h}_{ijk} \mathbf{v}_{\mathrm{i}_{jk}}^{\textrm{T}} \\&+ \sum_{\ell \in \mathcal{L}^{\mathrm{a}}} \sqrt{P_{\ell}} {\mathbf{g}_{i \ell} \mathbf{s}^{\mathrm{T}} } +
\mathbf{N}_i \Big) \mathbf{v}_{\mathrm{i}_{ik}}^{*}
\label{eqn:PC}
\end{aligned}
\end{equation}
where intra-cell pilot contamination is not present since BS $i$ allocates different pilots for different UEs in cell $i$.

\subsection{Precoder Calculation and Data Transmission}
\label{sec:precoder}

Thanks to the plurality of transmit antennas, 
BS $i$ is able to spatially multiplex $K_i$ UEs, 
while forcing $D_i$ nulls on the channel subspace $\mathrm{range}\{\widehat{\mathbf{\Sigma}}_i\}$ occupied by the neighboring Wi-Fi devices, as depicted in Fig.~\ref{fig:SystemModel}.

Let us define the estimated fast-fading channel matrix $\widehat{\mathbf{H}}_i \in \mathbb{C}^{N\times K_i}$ as
\begin{equation}
\widehat{\mathbf{H}}_i = \left[ \frac{\widehat{\mathbf{h}}_{ii1}}{\bar{h}_{ii1}},\ldots,\frac{\widehat{\mathbf{h}}_{ii{K_i}}}{\bar{h}_{ii{K_i}}} \right],
\label{eqn:Hi}
\end{equation}
obtained at BS $i$ normalizing the estimates $\widehat{\mathbf{h}}_{iik}$ in (\ref{eqn:PC}) by the slow fading channel component. 
Employing the normalized estimates in the precoder generally guarantees uniform UE average power allocation. 
The precoding matrix $\mathbf{W}_{ik}=[\mathbf{w}_{i1},\ldots,\mathbf{w}_{iK_i}]$ between BS $i$ and its served UEs is then obtained at every coherence interval as~\cite{Spencer04,GeraciJSAC,YanGerQueTSP2016}\footnote{
Note that zero forcing (ZF) precoding as used above has been shown to outperform maximum ratio transmission in terms of per-cell sum rate~\cite{BjoLarDeb2016}, 
and it can be extended to the case of multi-antenna UEs by considering block diagonalization~\cite{Spencer04}. 
When the system dimensions make the ZF matrix inversion in (\ref{eqn:precoder}) computationally expensive, 
a simpler truncated polynomial expansion can be employed with similar performance~\cite{KamMulBjo:2014}. 
Further improvements may be achieved by regularizing the inversion in (\ref{eqn:precoder}) \cite{Peel05,Geraci12,TatSmiGre2016}, or via interference alignment schemes~\cite{CadJaf2008,SuhHoTse2011,UstShaDmoICC2014}.}
\begin{align}
\mathbf{W}_{i} = \frac{1}{\sqrt{\zeta_i}} \widehat{\mathbf{H}}_i \left( \widehat{\mathbf{H}}_i^{\mathrm{H}} \widehat{\mathbf{H}}_i \right)^{-1},
\label{eqn:precoder}
\end{align}
where the constant $\zeta_i$ is chosen to normalize the average transmit power such that
\begin{equation}
\sum_{k\in\mathcal{K}_i}\Vert \mathbf{w}_{ik}\Vert^2 = 1.
\end{equation}
The precoder in (\ref{eqn:precoder}) employs the estimated channels obtained in (\ref{eqn:PC}) via projection on $\mathrm{null}\{\widehat{\mathbf{\Sigma}}_i\}$, and thus forces $D_i$ nulls on the channel subspace occupied by the neighboring Wi-Fi devices. We note that from a mathematical perspective, the projection onto $\mathrm{null}\{\widehat{\mathbf{\Sigma}}_i\}$ is equivalent to employing a virtual array with $N-D_i$ antennas. Due to the projection on $\mathrm{null}\{\widehat{\mathbf{\Sigma}}_i\}$, the matrix $\widehat{\mathbf{H}}_i$ has rank at most $\min\{K_i,N-D_i\}$. Therefore, the condition $D_i \leq N-K_i$ must hold for the inverse in (\ref{eqn:precoder}) to exist. Such condition is guaranteed by the inequality in (\ref{eqn:Di_fixed_Ki}).

The achievable rate at cellular UE $k$ in cell $i$ is given by
%\begin{equation}
%R_{ik}= \indic_{\mathbb{A}_i} \cdot \left( 1 - \frac{M_{\mathrm{c}}}{C_{\mathrm{c}}} - \frac{M_{\mathrm{p}}}{C_{\mathrm{p}}} - \frac{M_{\mathrm{LBT}}}{C_{\mathrm{TXOP}}} \right) \cdot \log_2 \left( 1 + \nu_{ik} \right)
%\label{eqn:Rik}
%\end{equation}
\begin{equation}
R_{ik}= \indic_{\mathbb{A}_i} \cdot \log_2 \left( 1 + \nu_{ik} \right)
\label{eqn:Rik}
\end{equation}
where the SINR $\nu_{ik}$ is given by (\ref{equ:sinr_ue}) using (\ref{eqn:precoder}) as the precoder, 
the notation $\indic$ denotes the indicator function, and $\mathbb{A}_i$ is the event of successful e-LBT operation defined in (\ref{eqn:LBT}).
%The factor between brackets in (\ref{eqn:Rik}) accounts for the overhead incurred by Wi-Fi channel covariance estimation, UE CSI acquisition, and e-LBT. The interval $C_{\mathrm{TXOP}}$ denotes the number of symbols in a transmission opportunity, after which a new e-LBT phase must be performed \cite{BiaTinSca2005}.
To avoid loss of generality by considering channel-specific parameters, 
in (\ref{eqn:Rik}), 
we have omitted a multiplicative factor accounting for the overhead incurred by Wi-Fi channel covariance estimation, UE CSI acquisition, and e-LBT. 
The expected cellular rate per BS $\bar{R}_{\mathrm{CELL}}$ is then obtained as
\begin{equation}
\bar{R}_{\mathrm{CELL}} = \frac{\mathbb{E} \left[ \sum_{i \in \mathcal{I}} \sum_{k\in\mathcal{K}_i} {R_{ik}} \right]}{\mathrm{card}\left\{\mathcal{I}\right\}} 
\label{eqn:rate_mean}
\end{equation}
where the expectation is taken with respect to all channel realizations and Wi-Fi traffic dynamics.
\section{Numerical Results}
\label{sec:simulations}

In this section, we evaluate the performance of the proposed mMIMO-U operations. 
We perform system-level simulations according to the scenario and methodologies described in Table~\ref{table:parameters}, 
unless otherwise specified. 
We first demonstrate the coexistence enhancement provided by mMIMO-U with respect to a conventional approach without Wi-Fi interference rejection. 
Then, we quantify the cellular data rates achievable in the unlicensed band. 
We also reveal the effect of an imperfect Wi-Fi channel covariance estimation. 
Finally, we discuss how the mMIMO-U spatial resources should be allocated as a trade-off between Wi-Fi interference suppression and cellular beamforming gain.

%802.11e standard defines default TXOP limit value for each AC, but values can be configured on AP. TXOP limit are set in intervals of 32µs (microseconds). Default TXOP is 47 for AC_VO (47×32=1504µs) for OFDM. It is 94 for AC_VI (94×32=3008µs).

\subsection{Enhanced Coexistence}

\begin{figure}[!t]
\centering
\includegraphics[width=1.015\columnwidth]{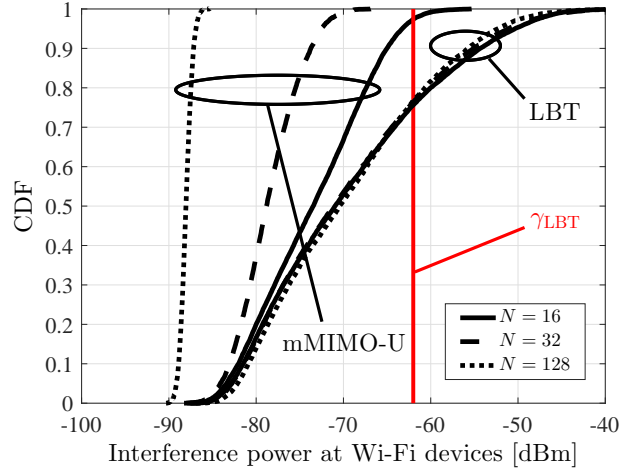}
\caption{Coexistence in the unlicensed band as seen by Wi-Fi devices.}
\label{fig:WiFiInt}
\end{figure}

\begin{figure}[!t]
\centering
\includegraphics[width=\columnwidth]{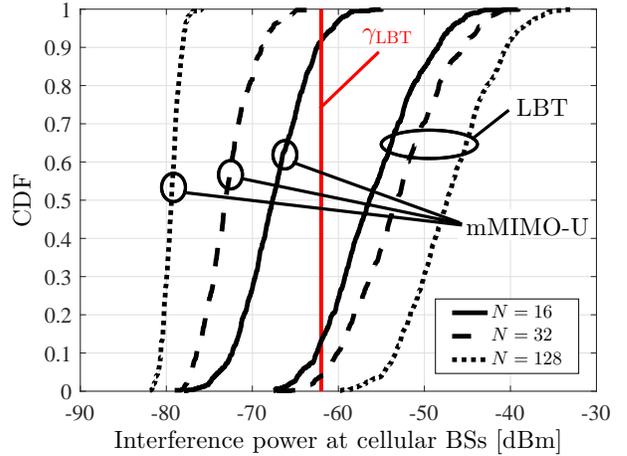}
\caption{Coexistence in the unlicensed band as seen by cellular BSs.}
\label{fig:BSInt}
\end{figure}

Figures~\ref{fig:WiFiInt}~and~\ref{fig:BSInt} show coexistence in the unlicensed band from the perspective of Wi-Fi devices and cellular BSs, respectively, 
comparing the proposed mMIMO-U to a conventional approach, 
where no Wi-Fi interference suppression is performed. 
The Wi-Fi channel covariance is computed via (\ref{eqn:Z}), 
and the behavior of both schemes is evaluated with an identical number of BS antennas.

Figure~\ref{fig:WiFiInt} shows coexistence in the unlicensed band from the perspective of Wi-Fi devices (both APs and STAs), 
assuming that cellular BSs have gained access to the unlicensed medium. 
The figure shows the cumulative distribution function (CDF) of the aggregate interference received by a Wi-Fi device, 
obtained from (\ref{eqn:Il}). 
With mMIMO-U, Wi-Fi devices are able to access the unlicensed band while BSs are transmitting. 
In fact, for $N\geq 32$, 
the aggregate interference is always below the regulatory threshold $\gamma_{\mathrm{LBT}} = -62$~dBm. 
On the other hand, with a conventional approach, 
Wi-Fi devices might not have access to the channel because the interference they receive is above $\gamma_{\mathrm{LBT}}$. 
Moreover, Fig.~\ref{fig:WiFiInt} shows that even when below the threshold, 
the aggregate interference received with a conventional approach is 50\% of the time above $-70$~dBm, 
which may affect the quality of Wi-Fi transmissions due to the non-negligible interference generated~\cite{jindal2015lte}. 
This phenomenon is not observed with mMIMO-U, 
as long as a sufficient number of antennas $N$ is available.

\begin{table}
\centering
\caption{Simulation parameters}
\label{table:parameters}
\begin{tabulary}{\columnwidth}{ |l | L | }
\hline
    \textbf{Parameter} 			& \textbf{Description} \\ \hline
  Cellular layout				& Hexagonal with wrap-around, 19 sites, 3~sectors each, 1 BS per sector \\ \hline
    Inter-site distance 		& 500m \cite{3GPP36814} \\ \hline
    UEs distribution 				& Random (P.P.P.), 32 UEs deployed per sector on average \\ \hline
	  UE association				& Based on slow fading gain \\ \hline
	UE pilot allocation			& Random with reuse 1 ($M_{\mathrm{p}} = 8$)  \\ \hline
	UE channel estimation  		&  Least-squares estimator \\ \hline
	Wi-Fi hotspots  & 2 outdoor hotspots per sector, radius: 20~m\ifx\[$\else\tablefootnote{We consider outdoor Wi-Fi devices since this case involves no wall penetration losses, making coexistence with cellular BSs more challenging.}\fi\\ \hline
	Wi-Fi devices  			& 8 devices per hotspot: 1~AP and 7~STAs \\ \hline
	Carrier frequency 		& 5.15 GHz (U-NII-1) \cite{FCC1430} \\ \hline
	%System bandwidth 		& 20 MHz (100 resource blocks with standard LTE parameters) \cite{FCC1430, 3GPP36814} \\ \hline
	System bandwidth 		& 20 MHz with 100 resource blocks \cite{FCC1430, 3GPP36814} \\ \hline
	Wi-Fi throughput  		& 65 Mbps per cluster \cite{PerSta2013}\\ \hline
	LBT regulations  		& Threshold $\gamma_{\mathrm{LBT}}=-62$ dBm \cite{PerSta2013} \\ \hline    
	d.o.f. allocation		& $K_i=8$ and $D_i = 0.5 (N-K_i)$ \\ \hline    
	%BS precoder 			& mMIMO-U: as in (\ref{eqn:precoder}), LBT: zero forcing \cite{Wagner12} \\ \hline
	BS precoder 			& As in (\ref{eqn:precoder}) \\ \hline
	BS antennas				& Downtilt: $12^{\circ}$, height: $25$~m  \cite{3GPP36814} \\ \hline
	%BS antenna height			&  \cite{3GPP36814} \\ \hline
	BS antenna array 		& Uniform linear, element spacing: $d = 0.5\lambda$ \\ \hline
	BS antenna pattern 		& Antennas with 3 dB beamwidth of $65^{\circ}$ and 8 dBi max. \cite{3GPP36873} \\ \hline
	BS tx power 			& 30 dBm \cite{FCC1430} \\ \hline   
	%Wi-Fi and UE antenna height 			& $1.5$ meters \cite{3GPP36814} \\ \hline
	Wi-Fi tx power 				& APs: 24~dBm, STAs: 18~dBm \cite{FCC2013}	\\ \hline   
	UE tx power		& Fractional uplink power control with $P_{0} = -58$ dBm and $\alpha = 0.6$ \cite{UbeVilRos2008} \\ \hline
	%Wi-Fi and UE antennas 		& 1 (omni-directional) \cite{3GPP36814} \\ \hline
	UE noise figure 			& 9 dB \cite{3GPP36824} \\ \hline
	UE rx sensitivity 			& -94 dBm \cite{3GPP36101} \\ \hline
    Fast fading  				& Ricean, distance-dependent K factor \cite{3GPP25996} \\ \hline
	Lognormal shadowing 		& BS to UE as per \cite{3GPP36814}, UE to UE as per \cite{3GPP36843} \\ \hline
	%Channel correlation			& Jakes correlation model \cite{656151} \\ \hline
	Path loss 					& 3GPP UMa \cite{3GPP36814} and 3GPP D2D \cite{3GPP36843}  \\ \hline
	Thermal noise 				& -174 dBm/Hz spectral density \\ \hline
\end{tabulary}
\end{table}

Figure~\ref{fig:BSInt} evaluates coexistence from the cellular BSs' standpoint, 
with mMIMO-U and with a conventional approach. 
It is assumed that Wi-Fi devices have gained access to the unlicensed medium, 
and the CDF of the interference received by cellular BSs is shown, 
obtained as the expectation of (\ref{eqn:LBT}) with respect to the symbols. 
Cellular BSs implementing the proposed mMIMO-U are generally able to access the unlicensed band, 
while Wi-Fi devices are transmitting. 
With $N = 16$ and $N = 32$ antennas, 
the aggregate interference received by the BSs is $90\%$ and $100\%$ of the time below the threshold $\gamma_{\mathrm{LBT}}$, respectively. 
On the other hand, BSs that perform conventional operations incur repeated backoffs, 
since their received interference is $87\%$ and $96\%$ of the time above $\gamma_{\mathrm{LBT}}$, respectively. 
Increasing the value of $N$ with the conventional approach yields a larger interference at these BSs, 
because more aggregate power is received. 
Instead, the proposed mMIMO-U drastically reduces such interference for increasing $N$, 
since an increasing number of d.o.f. are allocated for interference suppression.
%[David]: Note that the threshold at the mMIMO BS should be -72dBm and not -62dBm

\subsection{Achievable Data Rates}

\begin{figure}[!t]
\centering
\includegraphics[width=\columnwidth]{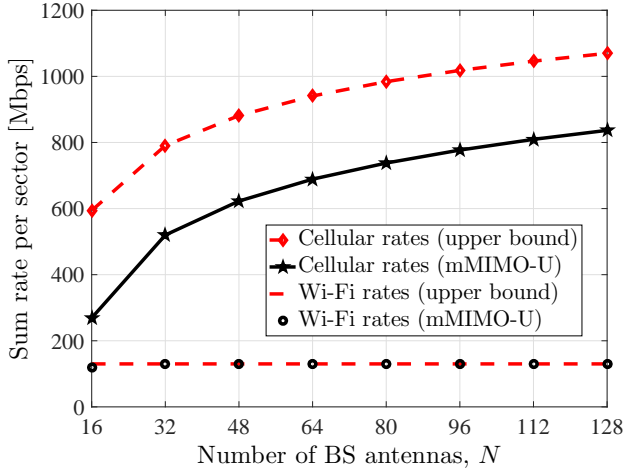}
\caption{Cellular and Wi-Fi data rates with proposed mMIMO-U versus number of BS antennas. Upper bounds on both rates are also shown, obtained in the ideal case of exclusive cellular and Wi-Fi use of the unlicensed band, respectively.}
\label{fig:Rates}
\end{figure}

\begin{figure}[!t]
\centering
\includegraphics[width=\columnwidth]{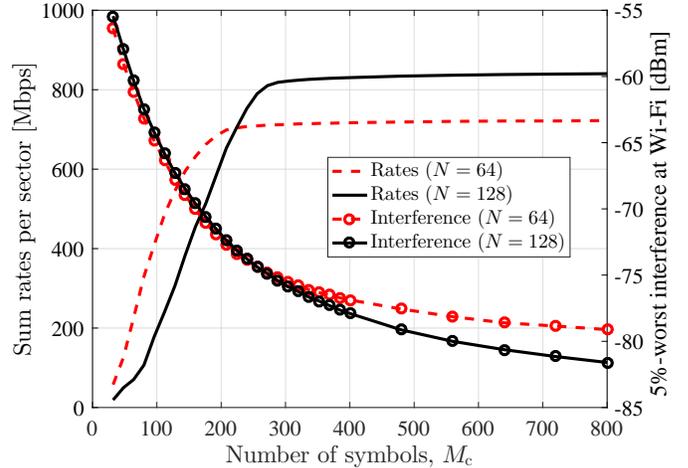}
\caption{Cellular mMIMO-U rates and interference generated at Wi-Fi devices versus number of symbols used for Wi-Fi covariance estimation.}
\label{fig:covariance}
\end{figure}

Figures~\ref{fig:Rates}~and~\ref{fig:covariance} show the data rates per cellular sector, obtained as in (\ref{eqn:rate_mean}). 
In Fig.~\ref{fig:Rates}, perfect knowledge of the channel covariance in (\ref{eqn:Z}) is assumed, whereas Fig.~\ref{fig:covariance} captures the effects of an imperfect covariance estimation. Moreover, note that Wi-Fi inter-cluster interference and collisions are neglected, Wi-Fi devices in a cluster are assumed active one at a time, and all rates provided by Wi-Fi APs are assumed equal to 65~Mbps when they gain access to the channel~\cite{PerSta2013}.\footnote{As discussed in Section~\ref{sec:conclusions}, a more accurate characterization of Wi-Fi rates in the presence of mMIMO-U transmissions requires higher-layer traffic models, e.g., those accounting for medium access control (MAC) protocols.}

Figure~\ref{fig:Rates} shows four curves: 
\emph{(i)} the Wi-Fi rates achievable with mMIMO-U; 
\emph{(ii)} the cellular rates achievable with mMIMO-U; 
\emph{(iii)} the Wi-Fi rates achievable when no cellular transmissions take place; and 
\emph{(iv)} the cellular rates achievable when no Wi-Fi transmissions take place. 
Note that \emph{(iii)} and \emph{(iv)} can be regarded as upper bounds for \emph{(i)} and \emph{(ii)}, respectively. 
The following observations can be made from Fig.~\ref{fig:Rates}. 
%[David]: In the next paragraph is not very clear how the indexing i), ii), iii) and iv) is used
First, for $N\geq 32$, the Wi-Fi rates achieved by mMIMO-U are constant across all values of $N$ and equal to the maximum value of 130~Mbps per sector. This reflects the fact that devices from both Wi-Fi clusters in the sector can access the medium 100\% of the time, since the received interference is always below $\gamma_{\mathrm{LBT}}$ as shown in Fig.~\ref{fig:WiFiInt}. 
Second, cellular rates with mMIMO-U are affected by the number of BS antennas $N$. For example, while 270~Mbps are achieved with $N=16$, cellular rates above 600~Mbps and 800~Mbps can be obtained by increasing $N$ to 48 and 112, respectively. In fact, as shown in Fig.~\ref{fig:WiFiInt}, a larger number of antennas also allows to suppress more interference to/from Wi-Fi devices, while leaving more spatial d.o.f. to multiplex cellular UEs with a larger array gain. Third, as the number of antennas $N$ grows, the gap between the cellular rates and the upper bound does not vanish since it is also due to the Wi-Fi-to-UE interference.

Figure~\ref{fig:covariance} draws the attention to two effects caused by inaccuracies in the Wi-Fi channel covariance estimate: degradation of the cellular rates and increased interference generated at Wi-Fi devices. To illustrate these phenomena, we show the achievable cellular rates and the 5th-percentile of $I_{: \rightarrow \ell}$ in (\ref{eqn:Il}), i.e., the 5\%-worst interference received by Wi-Fi devices during mMIMO-U operations. Both quantities are plotted versus the number of Wi-Fi samples $M_{\mathrm{c}}$ used to compute the estimate in (\ref{eqn:Z_hat}). The figure shows that as the number of samples $M_{\mathrm{c}}$ increases, the following occurs: the cellular rates grow, because the success rate of the e-LBT phase increases; and the interference at Wi-Fi devices diminishes, because the accuracy of the nulls increases. The value of $M_{\mathrm{c}}$ required to achieve large rates grows with $N$. Therefore, the Wi-Fi channel coherence interval poses a physical limitation to the number of BS antennas that can be effectively exploited~\cite{HoyHosTen2014}. 

\subsection{Spatial Resource Allocation}

\begin{figure}[!t]
\centering
\includegraphics[width=\columnwidth]{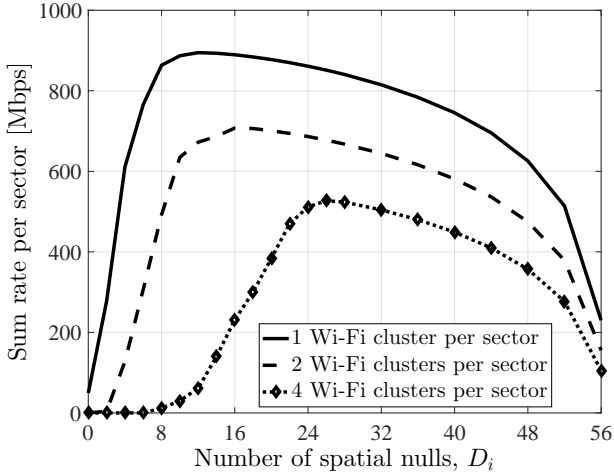}
\caption{Cellular mMIMO-U rates versus number of spatial nulls $D_i$ in the presence of one, two, and four Wi-Fi clusters per sector.}
\label{fig:Rates_vs_Di}
\end{figure}

\begin{figure}[!t]
\centering
\includegraphics[width=\columnwidth]{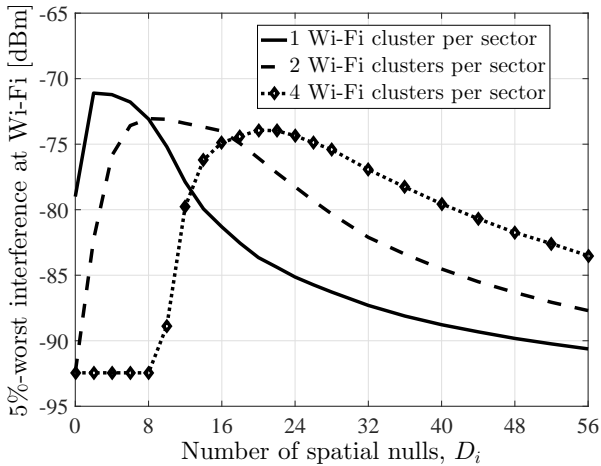}
\caption{Cellular-to-Wi-Fi interference versus number of spatial nulls $D_i$ in the presence of one, two, and four Wi-Fi clusters per sector.}
\label{fig:Interference_vs_Di}
\end{figure}

Figures~\ref{fig:Rates_vs_Di}~and~\ref{fig:Interference_vs_Di} illustrate the inherent trade-off between allocating more spatial d.o.f. for Wi-Fi interference suppression and employing them to augment cellular beamforming gain. 
In these figures, $N=64$ BS antennas and $K_i=8$ selected UEs per sector are considered. 
The number of spatial d.o.f. $D_i$ allocated for Wi-Fi interference suppression is varied to observe its impact. 
Three scenarios are considered, 
corresponding to one, two, and four Wi-Fi clusters per sector, respectively, with 8 Wi-Fi devices per cluster.

Figure~\ref{fig:Rates_vs_Di} shows the data rates per cellular sector as a function of $D_i$. Four observations are due: \emph{(i)} as $D_i$ increases from low values up to an optimal point, the rates increase because the e-LBT phase in (\ref{eqn:LBT}) is more likely to be successful; \emph{(ii)} as $D_i$ keeps increasing beyond the optimal value, the rates decrease because fewer d.o.f. are available for cellular beamforming gain; \emph{(iii)} the optimal value of $D_i$ increases with the number of Wi-Fi clusters per sector, 
because more nulls are required to suppress Wi-Fi interference; 
and \emph{(iv)} more Wi-Fi clusters correspond to lower cellular rates, 
because a larger Wi-Fi-to-UE interference is received.

Figure~\ref{fig:Interference_vs_Di} shows the 5\%-worst interference received by Wi-Fi devices. Similar observations can be made: \emph{(i)} as $D_i$ increases from low values up to a worst point, interference increases because more cellular BSs activate after successful e-LBT, thus more transmissions are generated; \emph{(ii)} as $D_i$ keeps increasing beyond the worst value, 
the interference decreases because more d.o.f. are employed to suppress it; 
\emph{(iii)} the optimal value of $D_i$ increases with the number of Wi-Fi clusters per sector, because more nulls should be employed to suppress Wi-Fi interference; and \emph{(iv)} for a given $D_i$, more Wi-Fi clusters correspond to larger interference, because Wi-Fi devices tend to occupy more spatial dimensions, out of which only $D_i$ can be nulled.

\section{Conclusion}
\label{sec:conclusions}

\subsection{Summary of Results}

We considered a mMIMO-U network, where massive MIMO cellular BSs and Wi-Fi devices operate in the same unlicensed band. We designed the main mMIMO-U scheduling and transmission operations to be performed at the BSs to enhance cellular/Wi-Fi coexistence. The scheduling procedures can be executed in a distributed fashion and include acquiring channel state information from the neighboring Wi-Fi devices, allocating spatial resources for Wi-Fi interference suppression and UE multiplexing, and selecting a suitable set of UEs to be served in the unlicensed band. For the transmission phase, we proposed to perform enhanced listen before talk, followed by UE pilot request, UE channel estimation, and precoder calculation. All along the mMIMO-U operations, the large number of BS antennas is exploited to suppress interference to/from neighboring Wi-Fi devices. As a result, Wi-Fi devices may access the unlicensed band as though no cellular transmissions were taking place, and vice versa. This enhances spatial reuse.

We evaluated the performance of mMIMO-U through simulations. Our results demonstrated the coexistence enhancement provided by mMIMO-U with respect to a conventional LAA-like approach. In fact, provided that cellular BSs are equipped with a sufficient number of antennas, mMIMO-U ensures that the mutual interference between cellular BSs and Wi-Fi devices falls below the regulatory threshold. We showed that large cellular data rates can be achieved without significantly degrading the performance of Wi-Fi networks deployed within the coverage area of a cellular BS. We finally discussed how the spatial resources made available by mMIMO-U should be allocated by compromising Wi-Fi interference suppression for cellular beamforming gain.

\subsection{Future Work and Discussion}

This work is suitable for several extensions from the system model, design, and deployment perspectives:
\begin{itemize}
\item 
\textit{Model:} 
Accurate traffic models are desirable for Wi-Fi devices and multiple operators sharing the same unlicensed band to evaluate how well BSs can estimate the channel covariance in (\ref{eqn:Z}). The rate computation at Wi-Fi devices should also account for these traffic models, since even when the received interference falls below the regulatory threshold, it may still affect the data rates~\cite{jindal2015lte}.
\item 
\textit{Design:} 
While in the current paper we focused on cellular downlink, appropriate procedures for mMIMO-U uplink should be defined. Coexistence between UE-to-BS transmissions and Wi-Fi transmissions must be guaranteed. 
One possible way to accomplish this would be to have BSs obtain access to the medium, and reserve it for their UEs to send uplink data in a synchronous manner.
\item 
\textit{Deployment:} 
An alternative strategy to the mMIMO-U scenario considered in this paper could consist of a more dense deployment of smaller low-power BSs, equipped with fewer antennas, covering smaller areas, and thus having to coexist with fewer Wi-Fi devices. Such deployment could allow, e.g., enterprise owners to roll-out high-performing indoor coverage without purchasing licensed spectrum from mobile network operators.
\end{itemize}

A final remark is due on emission regulations in unlicensed bands. Currently, in some countries, the maximum transmit power must be reduced for an increasing number of antennas, if the corresponding d.o.f. are used to focus energy in a particular direction \cite{FCC2013}. The scheme considered in this paper employs a large number of d.o.f. for UE multiplexing and Wi-Fi interference suppression, which led us not to account for the above guidance. Indeed, we expect future regulations to contemplate this aspect and consider adjustments to the guidance.

%\balance
\section*{Acknowledgment}

The authors would like to thank Dr.~Thomas~L.~Marzetta for his insightful comments.
\ifCLASSOPTIONcaptionsoff
  \newpage
\fi
\bibliographystyle{IEEEtran}
\bibliography{Strings_Gio,Bib_Gio}

\begin{thebibliography}{10}
\providecommand{\url}[1]{#1}
\csname url@samestyle\endcsname
\providecommand{\newblock}{\relax}
\providecommand{\bibinfo}[2]{#2}
\providecommand{\BIBentrySTDinterwordspacing}{\spaceskip=0pt\relax}
\providecommand{\BIBentryALTinterwordstretchfactor}{4}
\providecommand{\BIBentryALTinterwordspacing}{\spaceskip=\fontdimen2\font plus
\BIBentryALTinterwordstretchfactor\fontdimen3\font minus
  \fontdimen4\font\relax}
\providecommand{\BIBforeignlanguage}[2]{{%
\expandafter\ifx\csname l@#1\endcsname\relax
\typeout{** WARNING: IEEEtran.bst: No hyphenation pattern has been}%
\typeout{** loaded for the language `#1'. Using the pattern for}%
\typeout{** the default language instead.}%
\else
\language=\csname l@#1\endcsname
\fi
#2}}
\providecommand{\BIBdecl}{\relax}
\BIBdecl

\bibitem{GerGarLop2016ICC}
G.~Geraci, A.~{Garcia Rodriguez}, D.~L\'{o}pez-P\'{e}rez, A.~Bonfante,
  L.~{Galati Giordano}, and H.~Claussen, ``Enhancing coexistence in the
  unlicensed band with massive {MIMO},'' in \emph{Proc. IEEE Int. Conf. on
  Comm. (ICC)}, May 2017, accepted for publication.

\bibitem{Huawei2013}
{3GPP RP-131723}, ``Discussion paper on unlicensed spectrum integration to
  {IMT} systems,'' Dec. 2013.

\bibitem{QualcommMulteFire2015}
{Qualcomm News}, ``Introducing {MulteFire}: {LTE}-like performance with
  {Wi-Fi}-like simplicity,'' June 2015.

\bibitem{NokiaMulteFire2015}
{Nokia Executive Summary}, ``The {MulteFire} opportunity for unlicensed
  spectrum,'' 2015.

\bibitem{ZhaWanCai2015}
R.~Zhang, M.~Wang, L.~X. Cai, Z.~Zheng, X.~Shen, and L.~L. Xie,
  ``{LTE}-unlicensed: The future of spectrum aggregation for cellular
  networks,'' \emph{IEEE Wireless Communications}, vol.~22, no.~3, pp.
  150--159, June 2015.

\bibitem{ZhaChuGuo2015}
H.~Zhang, X.~Chu, W.~Guo, and S.~Wang, ``Coexistence of {Wi-Fi} and
  heterogeneous small cell networks sharing unlicensed spectrum,'' \emph{IEEE
  Comms. Mag.}, vol.~53, no.~3, pp. 158--164, Mar. 2015.

\bibitem{MukCheFal2015}
A.~Mukherjee, J.~F. Cheng, S.~Falahati, L.~Falconetti, A.~Furuskär, B.~Godana,
  D.~H. Kang, H.~Koorapaty, D.~Larsson, and Y.~Yang, ``System architecture and
  coexistence evaluation of licensed-assisted access {LTE} with {IEEE
  802.11},'' in \emph{{Proc. IEEE Int. Conf. on Comm. Workshop (ICCW)}}, June
  2015, pp. 2350--2355.

\bibitem{BenSimCzy2013}
M.~Bennis, M.~Simsek, A.~Czylwik, W.~Saad, S.~Valentin, and M.~Debbah, ``When
  cellular meets {WiFi} in wireless small cell networks,'' \emph{{IEEE} Commun.
  Mag.}, vol.~51, no.~6, pp. 44--50, June 2013.

\bibitem{PerSta2013}
E.~Perahia and R.~Stacey, \emph{Next Generation Wireless {LAN}s: 802.11n and
  802.11ac}.\hskip 1em plus 0.5em minus 0.4em\relax Cambridge University Press,
  June 2013.

\bibitem{LTEUForum:15}
{LTE-U Forum}, ``Coexistence study for {LTE-U} {SDL} {V1.0},'' \emph{{LTE-U}
  Technical Report}, Feb. 2015.

\bibitem{RahBehKoo2011}
M.~I. Rahman, A.~Behravant, H.~Koorapaty, J.~Sachs, and K.~Balachandran,
  ``License-exempt {LTE} systems for secondary spectrum usage: Scenarios and
  first assessment,'' in \emph{Proc. IEEE Sym. New Frontiers in Dynamic
  Spectrum Access Networks (DySPAN)}, May 2011, pp. 349--358.

\bibitem{Qualcomm2014}
{Qualcomm Research}, ``{LTE} in unlicensed spectrum: Harmonious coexistence
  with {Wi-Fi},'' June 2014.

\bibitem{3GPP36889}
{3GPP Technical Report 36.889}, ``Feasibility study on licensed-assisted access
  to unlicensed spectrum ({Release 13}),'' Jan. 2015.

\bibitem{RatManGho2014}
R.~Ratasuk, N.~Mangalvedhe, and A.~Ghosh, ``{LTE} in unlicensed spectrum using
  licensed-assisted access,'' in \emph{Proc. {IEEE} Global Telecomm. Conf.
  Workshops}, Dec. 2014, pp. 746--751.

\bibitem{3GPP-RP-140808}
{3GPP RP-140808}, ``Review of regulatory requirements for unlicensed
  spectrum,'' June 2014.

\bibitem{Nokia:14}
{Nokia}, ``Nokia {LTE} for unlicensed spectrum,'' \emph{{white paper}}, June
  2014.

\bibitem{CanLopCla2016}
C.~Cano, D.~Lopez-Perez, H.~Claussen, and D.~J. Leith, ``Using {LTE} in
  unlicensed bands: {P}otential benefits and co-existence issues,'' \emph{IEEE
  Commun. Mag.}, vol.~54, no.~12, pp. 116--123, Dec. 2016.

\bibitem{Mar:10}
T.~L. Marzetta, ``Noncooperative cellular wireless with unlimited numbers of
  base station antennas,'' \emph{{IEEE} Trans. Wireless Commun.}, vol.~9,
  no.~11, pp. 3590--3600, Nov. 2010.

\bibitem{RusPerBuoLar:2013}
F.~Rusek, D.~Persson, B.~K. Lau, E.~G. Larsson, T.~L. Marzetta, O.~Edfors, and
  F.~Tufvesson, ``Scaling up {MIMO}: Opportunities and challenges with very
  large arrays,'' \emph{IEEE Signal Process. Mag.}, vol.~30, no.~1, pp. 40--60,
  Oct. 2013.

\bibitem{LuLiSwi2014}
L.~Lu, G.~Y. Li, A.~L. Swindlehurst, A.~Ashikhmin, and R.~Zhang, ``An overview
  of massive {MIMO}: Benefits and challenges,'' \emph{{IEEE} J. Sel. Topics
  Signal Process.}, vol.~8, no.~5, pp. 742--758, Oct. 2014.

\bibitem{LarEdfTuf2014}
E.~G. Larsson, O.~Edfors, F.~Tufvesson, and T.~L. Marzetta, ``Massive {MIMO}
  for next generation wireless systems,'' \emph{IEEE Commun. Mag.}, vol.~52,
  no.~2, pp. 186--195, Feb. 2014.

\bibitem{BjoLarDeb2016}
E.~Bj\"{o}rnson, E.~G. Larsson, and M.~Debbah, ``Massive {MIMO} for maximal
  spectral efficiency: How many users and pilots should be allocated?''
  \emph{{IEEE} Trans. Wireless Commun.}, vol.~15, no.~2, pp. 1293--1308, Feb.
  2016.

\bibitem{MedTse2000}
M.~Medard and D.~N.~C. Tse, ``Spreading in block-fading channels,'' in
  \emph{Proc. Asilomar Conf. on Signals, Systems, and Computers}, vol.~2, Oct.
  2000, pp. 1598--1602.

\bibitem{HoyHosTen2014}
J.~Hoydis, K.~Hosseini, S.~T. Brink, and M.~Debbah, ``Making smart use of
  excess antennas: Massive {MIMO}, small cells, and {TDD},'' \emph{Bell Labs
  Tech. J.}, vol.~18, no.~2, pp. 5--21, Sept. 2013.

\bibitem{BjoSanDeb2016}
E.~Bj\"{o}rnson, L.~Sanguinetti, and M.~Debbah, ``Massive {MIMO} with imperfect
  channel covariance information,'' in \emph{Proc. Asilomar Conf. on Signals,
  Systems, and Computers}, Nov. 2016, to appear. Available as:
  \url{https://arxiv.org/pdf/1612.04128.pdf}.

\bibitem{ShiFosGan2000}
D.-S. Shiu, G.~J. Foschini, M.~J. Gans, and J.~M. Kahn, ``Fading correlation
  and its effect on the capacity of multielement antenna systems,''
  \emph{{IEEE} Trans. Commun.}, vol.~48, no.~3, pp. 502--513, Mar. 2000.

\bibitem{FCC2013}
{FCC 662911}, ``Emissions testing of transmitters with multiple outputs in the
  same band,'' Oct. 2013.

\bibitem{3GPP36201}
{3GPP Technical Specification 36.201}, ``{LTE}; {E}volved universal terrestrial
  radio access ({E-UTRA}); {LTE} physical layer ({R}elease 10),'' June 2011.

\bibitem{3GPP36213}
{3GPP Technical Specification 36.213}, ``{LTE}; {E}volved universal terrestrial
  radio access ({E-UTRA}); {P}hysical layer procedures ({R}elease 10),'' June
  2011.

\bibitem{ZhaSolLia2012}
J.~Zhang, P.~Soldati, Y.~Liang, L.~Zhang, and K.~Chen, ``Pathloss determination
  of uplink power control for {UL} {CoMP} in heterogeneous network,'' in
  \emph{Proc. {IEEE} Global Telecomm. Conf. Workshops}, Dec. 2012, pp.
  250--254.

\bibitem{3GPP36300}
{3GPP Technical Report 36.300}, ``Evolved universal terrestrial radio access
  ({E-UTRA}) and evolved universal terrestrial radio access network
  ({E-UTRAN}); {O}verall description ({R}elease 14),'' Sept. 2016.

\bibitem{JosAshMar2011}
J.~Jose, A.~Ashikhmin, T.~L. Marzetta, and S.~Vishwanath, ``Pilot contamination
  and precoding in multi-cell {TDD} systems,'' \emph{{IEEE} Trans. Wireless
  Commun.}, vol.~10, no.~8, pp. 2640--2651, Aug. 2011.

\bibitem{NgoLarMar2013}
H.~Q. Ngo, E.~G. Larsson, and T.~L. Marzetta, ``The multicell multiuser {MIMO}
  uplink with very large antenna arrays and a finite-dimensional channel,''
  \emph{{IEEE} Trans. Commun.}, vol.~61, no.~6, pp. 2350--2361, June 2013.

\bibitem{UbeVilRos2008}
C.~U. Castellanos, D.~L. Villa, C.~Rosa, K.~I. Pedersen, F.~D. Calabrese, P.~H.
  Michaelsen, and J.~Michel, ``Performance of uplink fractional power control
  in {UTRAN} {LTE},'' in \emph{Proc. IEEE Veh. Tech. Conference (VTC)}, May
  2008, pp. 2517--2521.

\bibitem{R1073224}
{R1-073224}, ``Way forward on power control of {PUSCH},'' in \emph{{3GPP TSGRAN
  WG1 49-bis}}, June 2007.

\bibitem{kay}
M.~Kay, \emph{{Fundamentals of Statistical Signal Processing: Detection
  Theory}}, P.~H. PTR, Ed., 1998.

\bibitem{Spencer04}
Q.~H. Spencer, A.~L. Swindlehurst, and M.~Haardt, ``Zero-forcing methods for
  downlink spatial multiplexing in multiuser {MIMO} channels,'' \emph{IEEE
  Trans. Signal Process.}, vol.~52, no.~2, pp. 461--471, Feb. 2004.

\bibitem{GeraciJSAC}
G.~Geraci, R.~Couillet, J.~Yuan, M.~Debbah, and I.~B. Collings, ``Large system
  analysis of linear precoding in {MISO} broadcast channels with confidential
  messages,'' \emph{IEEE J. Sel. Areas Commun.}, vol.~31, no.~9, pp.
  1660--1671, Sept. 2013.

\bibitem{YanGerQueTSP2016}
{H.~H.~Yang}, G.~Geraci, {T.~Q.~S.~Quek}, and {J.~G.~Andrews},
  ``Cell-edge-aware precoding for downlink massive {MIMO} cellular networks,''
  \emph{\emph{submitted to} IEEE Trans. Signal Process.}, 2016, available as
  arXiv:1607.01896.

\bibitem{KamMulBjo:2014}
A.~Kammoun, A.~M{\"u}ller, E.~Bj{\"o}rnson, and M.~Debbah, ``Linear precoding
  based on polynomial expansion: Large-scale multi-cell {MIMO} systems,''
  \emph{IEEE J. Sel. Topics Signal Process.}, vol.~8, no.~5, pp. 861--875, Jan.
  2014.

\bibitem{Peel05}
C.~B. Peel, B.~M. Hochwald, and A.~L. Swindlehurst, ``A vector-perturbation
  technique for near-capacity multiantenna multiuser communication - {P}art
  {I}: {C}hannel inversion and regularization,'' \emph{IEEE Trans. Commun.},
  vol.~53, no.~1, pp. 195--202, Jan. 2005.

\bibitem{Geraci12}
G.~Geraci, M.~Egan, J.~Yuan, A.~Razi, and I.~B. Collings, ``Secrecy sum-rates
  for multi-user {MIMO} regularized channel inversion precoding,'' \emph{IEEE
  Trans. Commun.}, vol.~60, no.~11, pp. 3472--3482, Nov. 2012.

\bibitem{TatSmiGre2016}
H.~Tataria, P.~J. Smith, L.~J. Greenstein, P.~A. Dmochowski, and M.~Shafi,
  ``Performance and analysis of downlink multiuser {MIMO} systems with
  regularized zero-forcing precoding in {R}icean fading channels,'' in
  \emph{Proc. IEEE Int. Conf. on Comm. (ICC)}, May 2016, pp. 1--7.

\bibitem{CadJaf2008}
V.~R. Cadambe and S.~A. Jafar, ``Interference alignment and degrees of freedom
  of the {K}-user interference channel,'' \emph{IEEE Trans. Inf. Theory},
  vol.~54, no.~8, pp. 3425--3441, Aug. 2008.

\bibitem{SuhHoTse2011}
C.~Suh, M.~Ho, and D.~N.~C. Tse, ``Downlink interference alignment,''
  \emph{{IEEE} Trans. Commun.}, vol.~59, no.~9, pp. 2616--2626, Sept. 2011.

\bibitem{UstShaDmoICC2014}
R.~F. Ustok, M.~Shafi, P.~A. Dmochowski, and P.~J. Smith, ``Interference
  alignment with combined receivers for heterogeneous networks,'' in
  \emph{Proc. IEEE Int. Conf. on Comm. (ICC)}, June 2014, pp. 5287--5292.

\bibitem{jindal2015lte}
N.~Jindal and D.~Breslin, ``{LTE} and {Wi-Fi} in unlicensed spectrum: A
  coexistence study,'' \emph{Google white paper}, 2015.

\bibitem{3GPP36814}
{3GPP Technical Report 36.814}, ``Further advancements for {E-UTRA} physical
  layer aspects ({R}elease 9),'' Mar. 2013.

\bibitem{FCC1430}
{FCC 14-30}, ``Revision of part 15 of the commission's rules to permit
  unlicensed national information infrastructure {(U-NII)} devices in the 5
  {GHz} band,'' Apr. 2014.

\bibitem{3GPP36873}
{3GPP Technical Report 36.873}, ``Study for {3D} channel model for {LTE}
  ({R}elease 12),'' June 2015.

\bibitem{3GPP36824}
{3GPP Technical Report 36.824}, ``{LTE} coverage enhancements ({R}elease 11),''
  June 2012.

\bibitem{3GPP36101}
{3GPP Technical Report 36.101}, ``User equipment ({UE}) radio transmission and
  reception ({R}elease 10),'' June 2011.

\bibitem{3GPP25996}
{3GPP Technical Report 25.996}, ``Spatial channel model for multiple input
  multiple output ({MIMO}) simulations ({R}elease 13),'' Dec. 2015.

\bibitem{3GPP36843}
{3GPP Technical Report 36.843}, ``Study on {LTE} device to device proximity
  services; radio aspects ({R}elease 12),'' Mar. 2014.

\end{thebibliography}
\balance
\end{document}